\documentclass[prl,twocolumn,showpacs,preprintnumbers,amsmath,amssymb,a4paper]{revtex4}
\usepackage{dcolumn}
\usepackage{bm}
\usepackage{graphics}

\begin{document}

\title{Natural multiparticle entanglement in a Fermi gas}

\author{Christian Lunkes$^1$  }
\email{christian.lunkes@imperial.ac.uk}
\author{ \v{C}aslav Brukner$^{2}$ }
\email{ caslav.brukner@univie.ac.at}
\author{ Vlatko Vedral$^{2,3}$ }
\email{ v.vedral@leeds.ac.uk} \affiliation{
$^1$QOLS, Blackett Laboratory, Imperial College London, London SW7 2BZ, England \\
$^2$Institut f\"{u}r Experimentalphysik, Universit\"{a}t Wien, Boltzmanngasse 5, A-1090 Vienna, Austria \\
$^3$The School of Physics and Astronomy, University of Leeds, Leeds LS2 9JT, England }

\date{\today}

\begin{abstract}
We investigate multipartite entanglement in a non-interacting fermion gas, as a
function of fermion separation, starting from the many particle fermion density
matrix. We prove that all multiparticle entanglement can be built only out of
two-fermion entanglement. Although from the Pauli exclusion principle we would always
expect entanglement to decrease with fermion distance, we surprisingly find the
opposite effect for certain fermion configurations. The von Neumann entropy is found
to be proportional to the volume for a large number of particles even when they are
arbitrarily close to each other. We will illustrate our results using different
configurations of two, three, and four fermions at zero temperature although all our
results can be applied to any temperature and any number of particles.
\end{abstract}


\maketitle                           

\textit{Introduction}. Quantum entanglement plays a crucial role in quantum mechanics,
and is extensively used in quantum information. However, it was only recently that
researchers have started to investigate entanglement in systems  containing a large
number of particles. This is of fundamental importance because entanglement was found
to  be relevant not only in microscopic systems, but also on a macroscopic scale
\cite{Caslav,Wang}. Multipartite entanglement seems to play an important role in
condensed matter systems, and might be the key ingredient to the solution of
unresolved physical problems such as high temperature superconductivity \cite{Vlatko}.
In this work we investigate multipartite entanglement in a non-interacting Fermi gas.
Bipartite entanglement in this simple quantum system has already been shown to be
fully characterized by the exchange integral due to the antisymmetry of the
wavefunction \cite{Vlatko2,Oh,Me}.

In this letter we  will show that all multiparticle entanglement can be built only
from bipartite entanglement. This is a significant result because it shows that a
complete description of quantum correlations at all levels is possible in a realistic
many body system such as a non-interacting Fermi gas. We show  that the $n$-particle
density matrix can be written as a sum of the completely mixed state and a mixture of
all  possible two fermion antisymmetrized wavefunctions. We use entanglement witnesses
to illustrate that genuine tripartite entanglement does not exist in this system, in
agreement with our previous expansion of the density matrix. We then investigate
bipartite entanglement for three and four  fermions for different fermion
configurations. This entanglement is  quantified using the negativity \cite{Vidal}.
Finally we show that for large number of fermions the entropy is always proportional
to this number (which in turn is proportional to the volume of the system),
independently of the fermion distance. For a small number of particles and small
fermion separation  the  entropy is smaller than this number (volume). This clearly
establishes the fact  that entanglement of a non-interacting Fermi gas can be treated
like any other macroscopic physical quantity and that it can be related to other
macroscopic observables such as the volume of the gas or number density. Any mean
field theory ignoring entanglement when describing macroscopic effects in many body
systems is therefore unlikely to be successful even when, remarkably, the constituents
of the system are non-interacting as in our case.

\textit{Density matrix}. We consider a many fermion system with a fixed number of
particles and a density matrix $\rho$. The elements of the reduced density matrices
for $1,2,3...n$ particles labeled by $\rho_{1}$, $\rho_{2}$, $\rho_{3}$...$\rho_{n}$
respectively are given by \cite{Yang}:
\begin{eqnarray}
\langle1|\rho_{1}|1'\rangle&=&\langle\Psi^{\dag}(1')\Psi(1)\rangle\nonumber\\
\langle12|\rho_{2}|1'2'\rangle&=&\langle\Psi^{\dag}(2')\Psi^{\dag}(1')\Psi(1)\Psi(2)\rangle\nonumber\\
\langle123|\rho_{3}|1'2'3'\rangle&=&\langle\Psi^{\dag}(3')\Psi^{\dag}(2')\Psi^{\dag}(1')\Psi(1)\Psi(2)\Psi(3)\rangle\nonumber\\
\langle1...n|\rho_{n}|1'...n'\rangle&=&\langle\Psi^{\dag}(n')\Psi^{\dag}((n-1)')...\Psi(n-1)\Psi(n)\rangle\nonumber
\end{eqnarray}
where $ 1\equiv(\textbf{r}_1,\sigma_1)$,  $\textbf{r}_1$ is the position vector and
$\sigma_1=\uparrow, \downarrow$ is the spin of the fermion. The average is given by
$\langle...\rangle=Tr\{\rho...\}$. For the sake of simplicity all our results are
illustrated  at  zero temperature, where $\rho=|\phi_0\rangle\langle\phi_0|$, with
$|\phi_0\rangle=\prod_k^{k_F}c^{\dagger}_{k,\sigma}|vac\rangle$ equals the ground
state of the Fermi system. The $c^{\dagger}_{k,\sigma}$ is the creation operator that
creates an electron of momentum $k$ and spin $\sigma$. The Fermi momentum is denoted
by $k_F$ and the vacuum state is $|vac\rangle$. The $\Psi$ are the field operators and
obey the usual fermion anti-commutation relations
$\{\Psi^{\dag}_{\sigma'_1}(\textbf{r}'_1),\Psi_{\sigma_1}(\textbf{r}_1)\}$
$=\delta_{\sigma'_1,\sigma_1}\delta(\textbf{r}_1-\textbf{r}'_1)$.

After a somewhat lengthy but straightforward calculation we arrive at a form for the
density matrix for $n$ particles which is particulary useful to investigate
entanglement:
\begin{equation}\label{eq:dm}
\rho_{n}=(1-\sum_{ij}p_{ij})\frac{\textbf{I}}{2^n}
+\sum_{ij}p_{ij}|\Psi_{ij}^{-}\rangle\langle\Psi_{ij}^{-}|\otimes
\frac{\textbf{I}}{2^{n-2}}
\end{equation}
where $|\Psi^{-}_{ij}\rangle
=1/\sqrt{2}(|\uparrow\downarrow\rangle-|\downarrow\uparrow\rangle)$ is the maximally
entangled singlet state of the pair $ij$. The sum runs over all the pairs $ij$. The
probabilities $p_{ij}$ are functions of  the relative distances between all pairs. As
an example, we write down the density matrix for the two and three particle case:
\begin{eqnarray}
\rho_{2}&=&p\frac{\textbf{I}}{4}+(1-p)|\Psi^{-}\rangle\langle\Psi^{-}|\nonumber\\
\rho_{3}&=&(1-p_{12}-p_{13}-p_{23})\frac{\textbf{I}}{8}
+p_{12}|\Psi_{12}^{-}\rangle\langle\Psi_{12}^{-}|\otimes \frac{\textbf{I}}{2}\nonumber\\
&+&p_{13}|\Psi_{13}^{-}\rangle\langle\Psi_{13}^{-}|\otimes\frac{\textbf{I}}{2}
+p_{23}|\Psi_{23}^{-}\rangle\langle\Psi_{23}^{-}|\otimes\frac{\textbf{I}}{2}\nonumber\\
\end{eqnarray}
 where $p=(2-2f(r)^2)/(2-f(r)^2)$
and $f(r)=j_1(x)/x$, with the Bessel function $j_1(x)=(\sin x- x \cos x)/x^2$ and
$x=k_F r$.  The relative distance between the fermion pair is denoted by $r$. The
function $f(r)$ is one for $r=0$  and zero for large $r$. For three fermions, we have
three different pairs and for the pair $ij$ :
$p_{ij}=(-f^2_{ij}+f_{ij}f_{ik}f_{jk})/(-2+f^2_{ij}+f^2_{ik}+f^2_{jk}-f_{ij}f_{ik}f_{jk})$.
The function $f_{ij}$ is a function of the relative distance between fermion $i$ and
$j$  only. Note that the probabilities $p_{ij}$ can be calculated for any number of
particles.

\textit{Entanglement}. The Peres-Horodecki criterion is the condition for the
existence of entanglement in the two particle case \cite{Peres}. In our earlier work,
we found this to imply that $f(r)^2>\frac{1}{2}$. This means that two electrons are
entangled if the relative distance between them is smaller then $1.8/k_F$ for $T=0$
\cite{Me}. Two fermions are maximally entangled if they are at the same position. This
is because of the Pauli exclusion principle. In general, since the overall state must
be antisymmetric, if the spatial  wavefunctions fully overlap and thus are symmetric,
then the spins must be antisymmetrised. The fermions must, therefore, be in the
maximally entangled spin singlet state $|\Psi^{-}\rangle$.
 We will show that such two-particle entanglement is also the main building block
 for multi-particle entanglement. In order to illustrate this behavior we will now consider
entanglement in systems containing three fermions .

 \textit{Tripartite entanglement}. From the decomposition of the density matrix it is clear that no
 genuine tripartite entanglement exists. We now formally show this using the method of entanglement witnesses.
 These are observables which (by our convention) have a positive expectation value for
all separable states, and a negative expectation value for some entangled states, i.e
entanglement exists if $Tr\{\rho^{(3)}\Pi\}<0$, where $\Pi$ is the witness
\cite{Toth}. It has been shown that there are only two different classes of tripartite
entanglement, which are represented by the
$|GHZ\rangle=1/\sqrt{2}(|\uparrow\uparrow\uparrow\rangle+|\downarrow\downarrow\downarrow\rangle)$
and
$|W_3\rangle=1/\sqrt{3}(|\uparrow\downarrow\uparrow\rangle+|\downarrow\uparrow\uparrow\rangle+|\uparrow\uparrow\downarrow\rangle)$
states \cite{Toth}. The corresponding witnesses are defined as:
$\Pi_{GHZ}=1/2-|GHZ\rangle\langle GHZ|$ and $\Pi_{W_3}=2/3-|W_3\rangle\langle W_3|$.
For both of these witnesses, because $0<|f_{ij}|^2<1$, the trace of $ \rho^{(3)}\Pi$
cannot be negative. This confirms  that genuine tripartite entanglement does not exist
in the ideal Fermi gas.

\textit{Bipartite entanglement }. We will now
 investigate if there is entanglement between two groups of
 fermions. One group contains fermion $i$ and the other group the  fermion pair $jk$.
 As a measure of entanglement we use the negativity, defined by $N_{[i,jk]}=(\|\rho_3^{T_i}\|_1-1)/2$, where $\|\rho_3^{T_i}\|_1$
is the trace norm of the partial transpose of the reduced density matrix $\rho_3$ of
fermion $i$ versus the other two $jk$ and denote it as $N_{[i,jk]}$.  The trace norm
can be evaluated to be
 $\|\rho_3^{T_i}\|_1=1+2|\sum_l\mu_l|$, where the sum
  goes over the negative eigenvalues of the partial
transpose. There are eight eigenvalues in total and only two of them are negative having
the same value $\lambda$. The negativity is, therefore, $N_{[i,jk]}=2|\lambda|$.
Negativity is a good measure of bipartite entanglement because it is monotonic under
local operations and classical communication and is also equal to zero if fermion $i$
is not entangled to the fermion pair $jk$. It reaches its maximal value of $1/2$  if
fermion $i$ is maximally entangled to the fermion pair $jk$. We will now investigate
different arrangements of three fermions, and investigate the behavior of negativity.

We first consider three fermions on a straight line.  The distance between the fermion
$i$ and the fermion $j$ is fixed. The remaining fermion $k$ moves from the position of
fermion $i$, to the position of  fermion $j$, i.e from  $x=0$ to $x_{max}=k_F r_{ij}$,
where $r_{ij}$ is the relative position between fermion $i$ and fermion $j$.  At
$x=0$, fermion $k$ is maximally entangled to fermion $i$, and can therefore not be
entangled to fermion $j$, independent of $r_{ij}$. The state of the total system is
then $\rho_{3}=|\Psi_{ik}^{-}\rangle\langle\Psi_{ik}^{-}|\otimes
\frac{\textbf{I}}{2}$. As fermion $k$  moves away from fermion $i$ we expect the
negativity $N_{[k,ij]}$ to first drop but and then to increase again as fermion $k$
approaches $j$. This is confirmed in Fig.1.
\begin{figure}
\begin{center}
\resizebox{8cm}{8cm}{\includegraphics{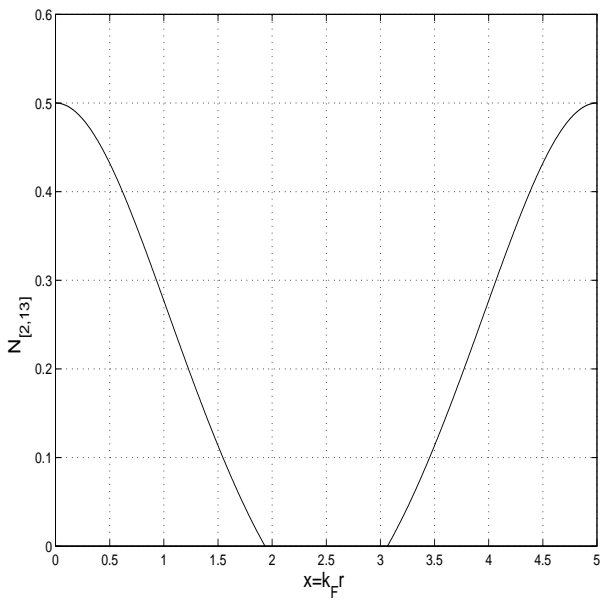}}
\end{center}
\caption{The negativity $N_{[2,13]}$ is plotted for fermion 2 moving from fermion 1 to
fermion 3. This corresponds to the text values $i=1$, $j=3$, $k=2$. The distance
between 1 and 3 is fixed to $x_{max}=5$. This is a dimensionless number.}
\end{figure}

We next consider the case where the fermions are located on the edges of an isosceles
triangle. Fermions $i$ and $j$ form the base of the triangle which is fixed. Fermion
$k$ is moved away from the midpoint of the base. The entanglement negativity
$N_{[k,ij]}$ and $N_{[i,jk]}$ for this scenario are plotted  in Fig.2. The
entanglement $N_{[k,ij]}$ monotonically decreases as $k$ moves away from $ij$ because
the effect of antisymmetrization becomes weaker with the distance (dashed line in Fig.
2). The entanglement negativity $N_{[i,jk]}$ (solid line in Fig.2) initially follows
the same trend as  $N_{[k,ij]}$ for exactly the same reason. Surprisingly however, the
entanglement $N_{[i,jk]}$, after reaching its minimum value, starts to increase and
then reaches its saturation value. The reason for this is the following. When fermion
$k$ is further away from $i$ and $j$ than the distance between fermion $i$ and $j$
itself, the effect of antisymmetrization between $i$ and $j$ on entanglement is larger
than the effect of antisymmetrization  between $i$ and $k$. If then the distance is
further increased, the position of fermion $k$ has a vanishingly small role on
entanglement. The three particle density matrix then becomes: $\rho_{3}\sim
\rho_{2}\otimes\textbf{I}/2$, where $\rho_{2}$ is the reduced density matrix of the
pair $ij$. Note that the minimum of negativity is reached when the fermions are
equally distant from each other. This is because entanglement is monogamous and each
particle has to share entanglement equally with the other two. This case will now be
analyzed in more detail.
\begin{figure}
\begin{center}
\resizebox{8cm}{8cm}{\includegraphics{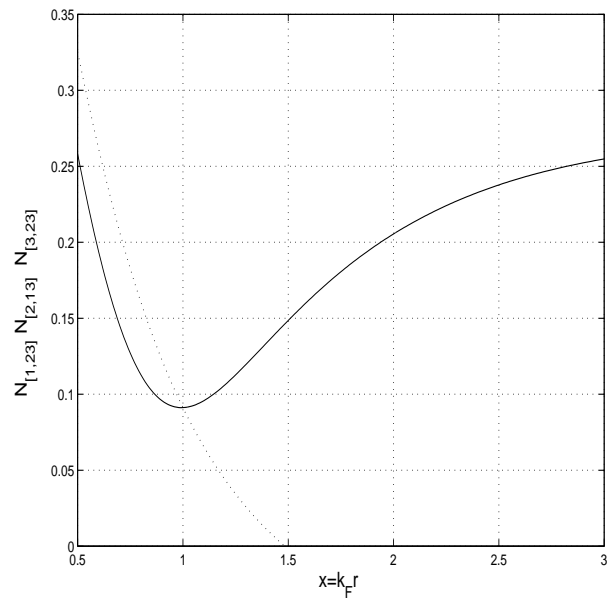}}
\end{center}
\caption{The negativities $N_{[2,13]}$ = $N_{[3,12]}$(solid), $N_{[1,23]}$ for an
isosceles triangle, for which fermions 2 and 3 form the base are plotted. Fermion 1 is
moving away from the midpoint of the base. This corresponds to the text values $i=2$,
$j=3$ and $k=1$. }
\end{figure}
\begin{figure}
\begin{center}
\resizebox{8cm}{8cm}{\includegraphics{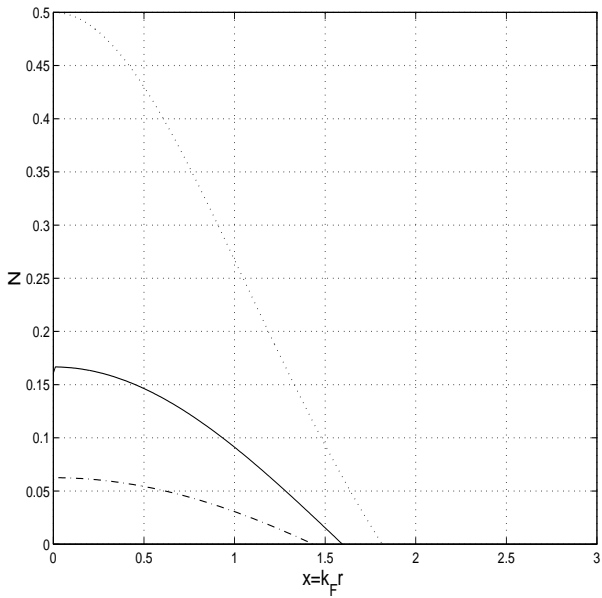}}
\end{center}
\caption{The negativities $N_{[1,2]}$, $N_{[1,23]}$(solid) and $N_{[1,234]}$(dashed)
are plotted. This corresponds to text values $p,q=1,2$, $l,mn=1,23$, and
$i,jkl=1,234$. }
\end{figure}

We now consider the case when the fermions are separated by equal distances. For three
particles the fermions are located on the edges of an equilateral triangle, and for
the four particle case on the edges of an tetrahedron. Entanglement in this case is
plotted Fig.3. We start by putting the fermions in a very small volume of radius
$\varepsilon$. Because of the Pauli exclusion principle, only two fermions can be in
the same location. More then two fermions would mean that at least  two quantum
numbers are the same, which is forbidden. As the distance between the fermions
increases, $N_{[i,jkl]}<N_{[l,mn]}<N_{[p,q]}$ at all distances, because the
entanglement is shared between the fermion pairs, and the more fermions are involved
the less entanglement we gain. All of this can be generalized to an arbitrary number
of fermions. If the fermions are all in a small volume  of radius $\varepsilon$, then
the state is in an equal mixture of all the singlet states of the pairs, and only the
second term of (\ref{eq:dm}) survives. Higher order entanglement does exist, but the
Pauli exclusion principle forbids maximally entangled states other then the
$|\Psi^{-}\rangle$. This is the reason why we do not have $|GHZ\rangle$ or
$|W_3\rangle$ states in the system. If the fermions are further away from each other
the first term in (\ref{eq:dm}) becomes important and the total state $\rho_n$ becomes
even less entangled. The interplay between the two terms in the density matrix is also
important when we want to calculate the total entropy of the fermions $S(n)=-Tr
\{\rho_{n}\ln \rho_{n}\}$. We study this quantity, because at $T=0$ it quantifies the
amount of entanglement between the measured electrons and the remaining unmeasured
electrons.

\textit{ Entropy}.
 We now investigate the von Neumann entropy of the Fermi system as
 a function of fermion distance as shown in Fig.4.  
 \begin{figure}
\begin{center}
\resizebox{8cm}{8cm}{\includegraphics{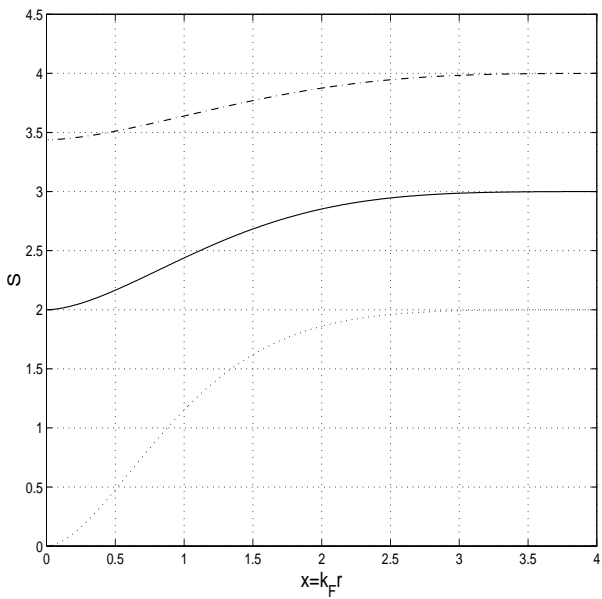}}
\end{center}
\caption{The von-Neumann entropy $S_{2}$(dashed), $S_{3}$(solid), $S_{4}$ is plotted
as a function of fermion distance. This behavior is analytically explained in the
text.}
\end{figure}
At $T=0$, the system is in a pure state. For the two particle case, the entropy is
zero for $x=0$ because the two fermions are in the pure singlet state. It then
increases to two as the distance increases. This behavior is also observed for three
particles as well as four. In these cases it does not start at zero  because we cannot
have more then two fermions in the same location, but again reaches the value of three
(four) as the
 distance increases. For large $n$, the entropy
becomes nearly equal to the number of particles even for very small distances. For a
large particle
 system at low constant density, the reduced density
matrix of the system is then given by $\rho_{n}=\textbf{I}/2^n$ and the entropy is
$S(n)=n \ln2$ which is proportional to the volume of the system. For a dense
 system the density matrix is given by:
$\rho_{n}=\sum_{ij}|\Psi_{ij}^{-}\rangle\langle\Psi_{ij}^{-}|\otimes
\textbf{I}/2^{n-2}$. The entropy of this state is also proportional to the number of
fermions for large $n$. Only if the number of fermions is small and if they are very
close, the Pauli exclusion principle prevents the entropy from being proportional to
the number of fermions. This can be explained as follows: If all the states are
equally likely the entropy is the logarithm of the number of all possible
configurations of these states. Since we have a system of fermions they have to be
antisymmetrized and therefore this number is equal to the total number of states
$2^n$, minus the number of symmetric states $(n+1)$ . The entropy therefore is:
\begin{equation}
S(n)=\ln (2^n-(n+1))
\end{equation}
We can see a complete agreement between this formula and the entropy in Fig. 4 for
$x=0$. It is also clear that if the number of fermions is large, then the first term
in the entropy dominates, giving us the previous result of entropy being proportional
to the volume of the system.

 \textit{Conclusion}. We have
presented a form of the density matrix for $n$ fermions in an ideal fermi gas which is
particulary useful to investigate entanglement in this system. We then showed that no
genuine multipartite entanglement exists, and that all multipartite entanglement can
be built only from the bipartite entanglement between fermion pairs. Lastly, we showed
that the entropy of a large Fermi gas is always proportional to its volume,
independently of fermion distance. It is only for a small number of fermions and small
distances that the  entropy is smaller than this number (volume). We believe that our
work shows that multipartite entanglement in complex macroscopic systems can be
studied and even fully understood with the existing techniques of quantum information.
We hope that this stimulates other studies in similar directions of solid state and
condensed matter systems.

\textit{Acknowledgments}: \v{C}. B. was supported by the Austrian Science Foundation
(FWF) Project SFB 1506 and by the European Commission (RAMBOQ). V. V. thanks European
Union and the Engineering and Physical Sciences Research Council for financial
support.


\begin{thebibliography}{9}
\bibitem{Caslav}\v{C}. Brukner and V. Vedral, e-print quant-ph/0406040 (2004).
\bibitem{Wang}X. Wang and P. Zanardi Phys. Lett. A \textbf{301},
1-2, (2002).
\bibitem{Vlatko}V. Vedral, New J. Phys. \textbf{6}, 22, (2004).
\bibitem{Vlatko2}V. Vedral, Central Eur. J. Phys. 1, 289-306,
(2003).
\bibitem{Oh}S. Oh and J. Kim, Phys. Rev. A \textbf{69}, 054305
(2004).
\bibitem{Me}C. Lunkes, \v{C}. Brukner and V. Vedral, to be published in Phys. Rev. A, e-print quant-ph/0410166 (2005).
\bibitem{Vidal}G. Vidal and R. F. Werner, Phys. Rev. A \textbf{65},
032314 (2002).
\bibitem{Yang}C.N. Yang, Rev. Mod. Phys. \textbf{34}, 4 (1962)
\bibitem{Peres}A. Peres, Phys. Lett. A \textbf{202}, 16 (1995); M.
Horodecki, P. Horodecki, and R. Horodecki, Phys. Lett. A \textbf{223}, 1 (1996).
\bibitem{Toth}G. T\'{o}th and O. G\"{u}hne, e-print quant-ph/0405165 (2004).
\end{thebibliography}
\end{document}